\documentclass[twocolumn]{article} 

\usepackage{graphicx}
\usepackage{hyperref}
\usepackage{twoopt}
\usepackage{authblk}
\newcommand{\ackname}{Acknowledgements}

\begin{document}
\title{Boosting the public engagement with astronomy through arts}

\author[1]{Valentin D. Ivanov}
\affil[1]{\small European Southern Observatory, Karl-Schwarzschild-Str. 2,
85748 Garching bei M\"unchen, Germany; vivanov@eso.org}
\date{2021-07-02}

\maketitle

\begin{abstract}
\begin{small}
Arts are a seamless way to introduce the general public to 
both basic and more sophisticated astronomical concepts. The 
visual richness of astronomy makes it attractive and easily 
incorporated in painting and literature. Astronomy is the 
only science with a muse -- Urania -- implying that, at least 
in the eyes of the ancients, it was an art itself. I review 
some less well known representation of astronomical concepts 
in literature with potential application in education.
\vspace{-5mm}
\end{small}
\end{abstract}
\vspace{1cm}

\section{Motivation}

The most ancient art is discovered on the walls of caves and
usually depicts everyday scenes -- somewhat direct studies of
the human condition in the immediate environments of its 
creators. The existence of such artistic examples is evidence 
that even in those much more difficult times than today, the 
humans have found resources to investigate the surrounding 
world, themselves and their place in that world via artistic 
tools. It is also a testament of the motivational power of art, 
needed to ensure those resources were allocated in epochs that 
we would call times of extreme scarcity. 

I set here a modest goal -- to use this power of art for 
educational purposes, taking advantage that it appeals to 
human emotions. The emotions are a key to long term memory 
\cite{Tyng_etal_2017}.
Furthermore, they provoke and feed human curiosity, posing 
with the same intensity the questions on different scales: 
what lies behind the neighboring hill and what lies behind 
the neighboring galaxy. 
Art equally provokes human imagination and charges the 
investigative spirit -- two necessary tools for addressing 
these questions.

Many educators have invoked art before in teaching to bring in
convenient examples and colorful illustrations of material 
that otherwise may seem to students unappealing and challenging 
at the same time. Michael Brotherton from the University of 
Wyoming in Laramie went even further and organized a series of 
{\it Launch Pad}\footnote{ \url{https://www.launchpadworkshop.org/}}
workshops to teach some basic astronomical concepts to 
interested writers.

Here I consider a few less well known literary works with 
potential application in astronomical education. They are 
dominated by the genre of hard science fiction, which by 
construction is ``science heavy''. Hopefully, they will be 
added to the astronomical teaching practices and will help both 
teachers and students. This summary is aimed at educators.

\section{Primers}

\subsection{Architecture of planetary systems in the novel 
{\it The Photon Starship}}

The discovery of the first hot jupiter 51\,Peg\,b 
\cite{1995Natur.378..355M} was a surprise for the astronomical 
community because our expectations for exoplanetary systems 
were based on the Solar system and it lacks an analogue of 
giant gas planets that move around their host stars on orbits 
tighter than that of Mercury. Later we discovered that the 
Solar system does not have representatives of another class of 
planets either -- the so-called super-Earths or mini-Neptunes, 
reported by \cite{2009ApJ...695.1006B} and 
\cite{2016ApJ...828...33D}, among others.

The system of 51\,Peg posed a difficult question how giant 
gas planets can exist so close to their star. Such planets 
could not have formed in situ because of the tidal forces. 
This leaves the orbital migration as a likely explanation 
\cite{1996Natur.380..606L}. It should be noted that the widely 
accepted Nice model of the Solar system evolution does include 
planetary migration (for a general description of the Nice 
model see \cite{2010CRPhy..11..651M}).

Exotic exoplanets are well covered in fiction. Probably, the 
most scientifically accurate and useful is the anthology {\it 
A Kepler’s Dozen} (2009), edited by S. Howell and D. Summers. 
It contains 13 stories, taking place on planets, discovered by 
the {\it Kepler} space telescope. However, the concept of an 
evolving planetary system is less well covered. 

A surprising example with a potential educational application 
was penned by Dr. Dimitar Peev (1919-1966), a science fiction 
writer and avid promoter of science from Bulgaria. His degree 
was in Law, but for the entire duration of his multi-decade 
career he worked as a science journalist, founding and 
contributing to the development of the two best known popular
science publications in his native country -- the magazine 
{\it Kosmos} and the newspaper {\it Orbita}.

\begin{figure}[!htb]
\centering
\includegraphics[width=8.1cm]{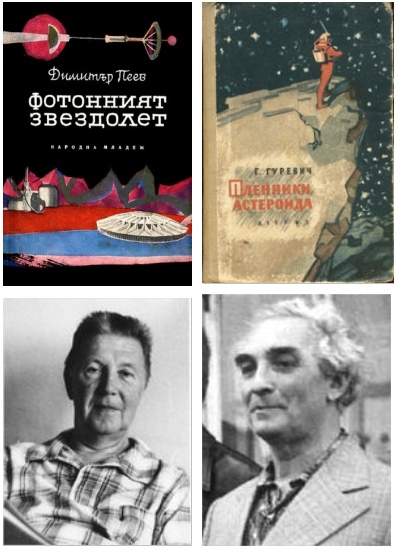}
\caption{Covers of the novel {\it The Photon Starship} and a 
collection {\it Prisoners of an Asteroid} that includes {\it
Infra Draconis} (top, left and right) and portraits of their 
authors Dr. Dimitar Peev and Georgi Gurevich (bottom, left and 
right).\vspace{-3mm}}\label{fig:bookcovers}
\end{figure}

Peev described in his novel {\it The Photon Starship} (1964;
Fig.\,\ref{fig:bookcovers}, left column) a crewed sub-light 
speed flight to our nearest neighbor in the constellation of 
Centaurus. The book follows the life of Aster, a kid born on 
the ship, who has never seen Earth. The story line includes the 
usual well-known surprises -- ``meteor attack, radiation attack, 
malfunctioning robotic kitchen''.

In the contest of this analysis, the most important element of
the novel is the discovery that the expedition makes upon 
arrival: Proxima is surrounded by a ultra-rich system of more 
than hundred planets with sizes between those of Mercury and 
Neptune. 

Here we should remind ourselves that Proxima, in addition to 
being the closest star to the Sun (at just over 4 light years), 
is the third component in a stellar system where the primary 
itself is a close stellar binary, known as Alpha Cen A and B. 
This is relevant for the analysis carried out by the fictional 
astronomers in the book. They find that the bodies orbiting
Proxima fall into two distinct classes:

(i) ``Native'' planets, with orbits nearly aligned into a 
single plane, similar to the planets in the Solar system. The 
scientists in the book speculate that these have been formed
in the protoplanetary disk that once have surrounded Proxima 
itself, much like the planet of our system have formed in the 
protoplanetary Solar disk.

(ii) Accreated planets on highly elliptical orbits, 
with random orbital inclinations. These 
planets are likely descendants of captured planetesimals that 
have been ejected from the neighboring Alpha Cen A and B early 
in the history of that system. The stability of this system is
not discussed.

The book offers background material for explaining a number of
astronomical concepts. The stellar multiplicity and the various
mechanisms of binary formation are perhaps the most obvious 
ones. A suitable topic for discussion in class is the validity 
of the fictional hypothesis, explaining the existence of the 
accreted part of the Proxima's planetary system: is it 
realistic in the light of the more recent -- with respect to 
the publication year of the novel -- discovery that the chemical 
composition of Alpha Cen A and B is different than that of 
Proxima? This was found during detailed abundance analysis of 
the three stars: [Fe/H]=0.22$\pm$0.10 to 0.26$\pm$0.08 
\cite{2015A&A...582A..81J} for Alpha Cen A and B, and 
[Fe/H]=$-$0.07$\pm$0.14 \cite{2016A&A...587A..19P} to 
0.05$\pm$0.20 \cite{2016A&A...593A.127K} for Proxima (the 
statistical significance of that difference is yet another 
topic for discussion that I will only point at here). This 
implies that 
while the central pair has formed together, while Proxima is a 
later acquisition, most likely being accreted during a stellar 
flyby. The time of this flyby is unknown, but the probability 
that it might have happened early enough while the central 
binary still had multiple planetesimals is low, just because 
this is a short period -- of up to a few tens of millions of 
years 
\cite{2016MNRAS.461.2257K,2020MNRAS.492.3849K} -- with respect 
to the billion year age of the two main stars in the system. 
Dynamic studies of the Alpha Cen system also favor different 
origin of its components \cite{2018MNRAS.473.3185F}.

Stellar flybys like the one during which Proxima joined the 
Alpha Cen system are extremely rare on human lifetimes, but 
common by astronomical standards. Our Solar system underwent 
one only about 70 thousands of years ago 
\cite{2015ApJ...800L..17M,2015AJ....149..104B,2015A&A...574A..64I}, 
but the passing star had only $\sim$15\,\% of the Solar mass 
(Proxima is comparable -- 12\,\% of the Solar mass). Such 
events rise the question of the fragility of the planetary 
systems \cite{2015MNRAS.451..144P}. They are difficult to 
detect, because the encounters are statistically likely to 
involve the most common type of stars in the Milky way -- 
low-mass, cool and intrinsically faint objects -- the reason 
why Proxima itself was discovered only in 1915 by Robert 
Innes \cite{1915CiUO...30..235I,1917CiUO...40..331I} when 
astronomical photography became a common tool, making it 
possible to carry out studies of proper motions of thousands of 
stars. The work of Innes was an early but important precursor 
of moderns space missions like HIPPARCOS 
\cite{1997A&A...323L..49P} and Gaia \cite{2016A&A...595A...1G}
that draw detailed maps of the Milky Way and its nearest 
neighboring galaxies.


One can use the novel as a topic starter for a discussion on
exoplanetary systems' architecture as well. If we set aside 
the Solar system (keeping in mind the argument about the status 
of Pluto\footnote{ A non-fictional account of the Pluto debate
can be found in {\it How I Killed Pluto and Why It Had It 
Coming}, by Michael E. Brown, 2010.} and about the hypothetical 
Planet IX -- see \cite{2016AJ....151...22B,2017NewSc.233R..24B}), 
then the 
richest planetary system we know of is Kepler-90, with eight 
planets \cite{2018AJ....155...94S}. Incidentally, Kepler-90 is 
a G type solar analog, but the majority of richest planetary 
systems are found around M stars, just like Proxima. An example 
of that is the seven-planet TRAPPIST-1 
\cite{2017Natur.542..456G}.
This apparent preference for late type host stars reflects the 
observational fact that low mass stars tend to be orbited by 
more, but lower mass planets, in comparison with Solar type 
stars \cite{2013ApJ...767...95D}, although some planets may 
have evaded detection \cite{2021arXiv211208337B}. Indeed, 
\cite{2006tafp.conf..111B} noticed some time ago that hot 
Jupiters are rarely found around cool low-mass stars. Later 
on this was explained with the limitations of the planet 
formation in the protoplanetary disks -- it is easy to 
understand this intuitively: the lower the stellar mass, the 
lower the mass of the protoplanetary disk, the lower the mass 
of the planets that can form inside them. The effect is 
boosted by an additional bias that the lower mass planets 
are easier to discover around lower mass stars, than around 
higher mass stars 
\cite{2020A&A...633A.116M,2021MNRAS.tmp.3293P}.

It is unlikely that Dimitar Peev predicted by a stroke of 
genius the fine details of the planetary formation in the 
vicinity of M stars. Still, the agreement between the 
properties of his fictional planetary system and the real 
planetary systems of M stars is not entirely a chance 
coincidence, because in an attempt to make the his fictional 
world as realistic as possible, he based the reality of the
novel on first physical principles. A number of important 
phenomena were taken into account: the scattering of 
protoplanets by the two close-in companions, the conservation 
of momentum that would preserve the orientation of planetary 
orbits in the plane of the protoplanetary disk. There are 
some inconsistencies or rather improbable elements, but they 
could only be recognized based on information that was not 
available at the time when the book was written.

\subsection{Pre-discovery brown dwarfs in the novella 
{\it Infra Draconis}}

The novella {\it Infra Draconis} was first serialized in a
popular science magazine {\it Knowledge -- power} in late-1958. 
It was written by Georgi 
Gurevich (1917-1998; Fig.\,\ref{fig:bookcovers}, right
column) -- a Soviet and Russian science fiction writer. 
Interestingly, this novella has appeared in English twice, in 
collections of science fiction stories by Soviet writers: {\it 
A Visitor from Outer Space} (1961, Foreign Language Press, 
Moskow) and {\it Soviet Science Fiction} (1962, Collier, New 
York). The latter has an introduction by Isaac Asimov.
Before becoming a full-time writer Gurevich had a successful 
career as a construction engineer. He also published many 
popular science articles in newspapers and magazines -- which 
seems to be a common pattern among the writers who produced 
works suitable for educational purposes.

The story follows a space flight to an ultra-cool object, 
called infra, located in the Draco constellation. We learn 
that in the fictional world these are recently discovered 
objects, too small to sustain fusion in their cores. The 
infras are analogues of the real-world brown dwarfs (BDs). 

Hypothetical dark substellar objects floating freely in space 
and not massive enough to sustain hydrogen fusion were first 
considered in the scientific literature only five years after
the publication of {\it Infra Draconis} by 
\cite{1962AJ.....67S.579K} who named them black dwarfs. They 
still burn Lithium, though, unlike planets -- their even 
lower mass ``cousins''. The name was later replaced with 
``brown dwarf'', as a truer description of their colors, as 
suggested by Jill Tarter, better known for her SETI work. The 
first BD was observed in 1988 -- it is a companion to the 
white dwarf GD165, but it was not recognized that it belongs 
to the class for a decade \cite{1997A&A...327L..29M}. Modern 
theoretical models suggest these objects span a range between 
$\sim$12 and $\sim$60\,M$_{Jup}$
\cite{2013MSAIS..24..128A,2014ApJ...787...78M}.

It is too late to ask the late Georgi Gurevich how he came up
with the idea for sub-stellar objects. In a similar situation, 
Frederick Pohl was asked how he came up with the idea that a 
super-massive object resides at the center of every galaxy --
a statement made by one of the characters in his novella ``The 
Gold at the Starbow's End'' (1972). Apparently, he read it in 
a popular science journal and with some speculation this 
hypothesis -- at the time -- can be traced to works by Salpeter 
and Zeldovich published at around the same time
\cite{1964ApJ...140..796S,1965SPhD....9..834Z}

The closest infra in the fictional world of Gurevich is only 
seven light days away -- a much smaller distance than to the 
nearest star Alpha Cen (discussed in the previous section). 
This is also the most interesting infra, because its surface 
temperature is 10\,C -- about the average winter day-time 
temperature on the Bulgarian Black Sea coast (at about 42\,deg 
north of the equator). This puts Infra Draconis at the 
temperature range allowing existence of liquid water. Unlike 
Earth, with its significant diurnal-nocturnal (day-to-night) 
temperature 
variations, the infras have constant surface temperatures, 
because they are heated internally. Their energy source is 
contraction -- for the same reason Jupiter emits more energy 
than the amount it receives from the Sun.

The kosmonavts from the novella spend fourteen years to reach 
their target, but in the middle of the journey they discover 
that the Earth-side observations were wrong -- Infra Draconis 
is binary object. The two components have surface temperatures 
of $-$6\,C and 24\,C. The original planning sends the 
expedition to the cooler object -- a giant methane and ammonia 
gas ball, similar to Jupiter in our Solar system. However, the 
warmer one turns out to be a liquid water world. The 
expedition's leader makes the ultimate sacrifice, descending 
below the surface (captains going on away missions is not a 
tradition limited to Starfleet), only to discover in his last 
moments a thriving technological extraterrestrial civilization 
(ETC) at the bottom of the alien ocean.

In the real world the nearest binary BD we know of is Luh-16 
(a.k.a. WISE\,J104915.57$-$531906.1), located at about 2\,pc 
from the Sun \cite{2013ApJ...767L...1L}, but the effective 
temperatures of both its components place them well above the 
liquid water regimen. Much closer analog of the fictional 
Infra Draconis is the BD WISE\,J085510.83$-$071442.5 (often 
abbreviated to W0855) -- as far as we know, a single object -- 
with an effective temperature $\sim$250\,K (close to $-$20\,C) 
at a distance of 2.2\,pc from the Sun
\cite{2014ApJ...786L..18L,2014A&A...570L...8B}. It is probably 
3--10 times more massive than Jupiter and its age range is 
poorly constrained -- it may lay anywhere between 1 and 10\,Gyr.

Why were these nearby objects discovered only recently? In fact, 
Luh-16 was detected in a number of surveys going as far back as 
the 1970-90s \cite{2013arXiv1303.5345M}, but its high proper 
motion (8.1$\pm$0.1\,arcsec\,yr$^{-1}$) and the crowded region 
of the sky it resided in prevented astronomers from recognizing 
that these detections belong to the same object. The 
non-detection of W0855 is due to its intrinsic faintness 
\cite{2014A&A...570L...8B,2014ApJ...797....3K,2016ApJ...152...78L,2016A&A...592A..80Z}.

Life on planets around BDs was considered by 
\cite{2020ApJ...888..102L} (in fact, the first directly imaged 
planet orbits a brown dwarf 
\cite{2004A&A...425L..29C,2005A&A...438L..25C}. They concluded 
that only more massive BDs (M$\geq$30\,M$_{Jup}$) are likely to 
sustain planets with habitable conditions on sufficiently long 
time scales. Surprisingly, the idea of Gurevich that life may 
thrive inside the atmospheres of a BD has also been studied by 
\cite{2017ApJ...836..184Y} and \cite{2019ApJ...883..143L}. They 
argue that there are a few tens of W0855like objects within ten 
parsecs from the Sun, and that the habitable volume in them may 
be hundred times larger than on Earth-like planets.

In the context of this summary the novella {\it Infra Draconis} 
is suitable to introduce the ultra-cool sub-stellar mass objects, 
to underline their high spatial density, and the possibility that
they may harbor life -- either on planets or within their own
atmospheres. The latter topic may be particularly appealing to 
the public: recent data suggests that our Solar system may harbor 
up to ten subsurface oceans, compared with only one planet with a 
habitable surface -- Earth. If the subsurface oceans are viable
habitable zones, this may increase the habitable space of a 
planetary system by an order of magnitude.

\subsection{Astronomical methods and social environment in the 
astronomical community in the novel {\it Stars and Waves}}

The novel 
{\it Stars and Waves}\footnote{ \url{https://www.starsandwaves.net/}}
was written by the well-known astronomer Roberto Maiolino 
from University of Cambridge. He is specializing in 
extragalactic studies and observational cosmology. This book 
is different from previous examples. First, because it is a 
work of a professional astronomer who knows the research 
process and astronomical environment inside out. Second, 
because it is marketed as a thriller. 

{\it Stars and Waves} offers information about various aspects 
of exoplanet research and paints a realistic and detailed view 
of the life in the astronomical community at a page-turning 
pace. An important disclaimer: in the real world the 
astronomers' lives do not lead to untimely ends as a result of 
fierce competition to publish, and generally, the life in 
academia is not amply sprinkled with violence, as the readers 
may be led to believe -- these are only plot drivers and 
concessions to the genre. The book can be used as educational 
tool, but for more mature audience. 

The novel often detours into the history of astronomy -- the 
work of Newton, Galileo and others are prominently featured in 
an inspirational manner. Some relevant past and present-day 
technologies are also explained, so the novel can provide 
background for a discussion how fast the technological progress 
enables the astronomical progress.

The central scientific question in {\it Stars and Waves} -- not
revealed at the beginning -- is of finding life elsewhere in the 
Universe. The answer had been pursued in two ways. The typical 
programs to search for advanced technological extraterrestrial 
intelligence (SETI) usually rely on various leaked emission -- 
for example, radio from communications or heat from massive 
computation or from Dyson spheres 
\cite{1966ApJ...144.1216S,1973Icar...19..350S,2021AcAau.188..203W}.
We argued in \cite{2020A&A...639A..94I}  
that advanced technological ETCs are likely to reduce such 
losses, rendering these strategies inefficient. This leaves 
us with the options of communicating either with advanced ETCs
that are willing to set up beacons intentionally aimed at and 
optimized for ``younger'' cousins like us, or with nearby (in 
both space and development) peer-level ETCs that allow 
significant radio or thermal leaks. 

The second path to search for life in the Universe is to look 
for traces of the natural products of this life -- the so-called
biosignatures \cite{1981Icar...46..293N,2005AsBio...5..706S}.
An excellent introductory reading material can be found at the 
NASA Astrobiology 
page\footnote{ \url https://astrobiology.nasa.gov/education/alp/what-is-a-biosignature/}.
The Oxygen and Ozone are the most informative biosignatures.
Strictly speaking, water is not a biosignature, but it is an 
important solvent and its presence is considered critical for 
life as we know it. The Carbon dioxide and Methane are other 
potential indicators of life, although some geological sources 
can contribute to their content in the atmospheres of exoplanets 
\cite{2002AsBio...2..153D,2018AsBio..18..663S}.

{\it Stars and Waves} presents to the reader an intriguing 
combination of bio- and technosignature based ETC searches.

On more technical level, Maiolino introduces the transmission 
spectroscopy as a main exoplanet characterization tool -- in 
the real world this is a commonly used technique that brought 
up the first direct detections of exoatmospheres 
\cite{2002ApJ...568..377C,2003Natur.422..143V}. 
This method can be traced further back to the Russian scientist 
Mikhail Lomonosov (1711-1765) who discovered the atmosphere of 
Venus during a transit of that planet across the Sun in 1761.
\cite{2018haex.bookE.100K} summarizes the present-day status of 
the transmission spectroscopy.

The novel is written in what almost qualifies for a 
semi-documentary style, the action takes place on existing (and 
exotic) 
astronomical locations and it draws quite realistic picture of 
every-day's life of researchers and astronomical institutions. 
Some concessions to action have been made, though: most large, 
modern observatories maintain stringent account for every 
second of observing time, or every obtained image or a spectrum, 
especially if they are funded by public money. Unsanctioned 
target changes at such observatories are hardly ever possible. 
Finally, the book 
spectacularly succeeds conveying the competitive academic 
atmosphere.

{\it Stars and Waves} can be used to stimulate discussion and 
interest towards a number of subjects, ranging from astronomical 
instrumentation, including spectroscopy; concepts why a certain 
site is suitable or not for observations, including dryness and 
light pollution. A rarely touched upon topic is how actually the 
work at observatories is organized and the novel invites a
comparison with the Commencement address that the famous 
astronomer Edward Pickering (1846-1919) gave at Case School of 
Applied Science in Cleveland on May 27, 1909. He outlined his 
uniquely insightful view of astronomical institutions of the 
future, and they look surprisingly similar to the modern 
observatories that Maiolino describes.


\section{Summary}

I describe a few literary works that can be useful for 
astronomical education. They help to introduce many basic 
concepts, including binary stars, planet formation, brown 
dwarfs. The big question about the search for other life in
the Universe is also covered.

This material can be used in different manners -- as 
introductory points, as basis for class discussion, or as 
cases to be troubleshooted by the students -- and for 
students at different level, from high school to amateur
astronomers \cite{2018arXiv181201582I} and even graduate 
level students who can exercise debugging some of the false 
concepts and conclusions that were present in these works.

A number of novels and stories that can be useful to 
educators were omitted. Among them are the anthologies
{\it Diamond in the Sky} (2008) and {\it Science Fiction by 
Scientists} (2017; both edited by M. Brotherton), the novel 
{\it Les Robinsons du Cosmos} ((The Robinsons of the Cosmos; 
1955) by the French SciFi writer Francis Carsac (1919-1981), 
the upcoming astrobiological anthology {\it Life Beyond Us} 
(2022, Eds. S. Forest, L. K. Law, J. Nov{\'a}kov{\'a}), and 
in particular the excellent non-fiction teaching manual {\it 
Exploring Science Through Science Fiction} (2013) by B. B. 
Luokkala.

Finally, this paper offers an eclectic mix of references, 
ranging from fiction, to popular articles and research works 
to help educators in finding sources that match the level of 
their audience.

\section*{\ackname}
This is an extended write up of a poster presented at the European 
Astronomical Society (EAS) Annual Meeting held on line, Jun 28 -- 
Jul 2, 2021, Special Session SS6 (Jul 2, 2021): Diversity and 
Inclusion Day. I thank the organizers for giving me the opportunity 
to demonstrate the power of art as an educational tool.

{\small

}

\end{document}